\documentclass[a4paper,UKenglish,cleveref, autoref, thm-restate]{oasics-v2021}
\pdfoutput=1

\usepackage{listings}
\definecolor{lightgray}{RGB}{225,225,225}

\lstdefinelanguage{whiley}
{
        keywords={import, function, method, property, type, constant, assert, assume, for, while, switch, is, if, case, return,
          else, process, define, as, requires, ensures, where, no, some, new,
          all, bool, int, byte, char, string, void, real, in, any,
          null, var, public, protected, private, skip, break, do,
          throws, catch, continue, default, try, fail, export, final
        },
        basicstyle=\ttfamily, 
        commentstyle=\small\rmfamily\itshape,
        stringstyle=\small\itshape,
        moredelim=*[s][commentstyle]{/*}{*/}, 
        morecomment=[l][commentstyle]{//},      
        backgroundcolor=\color{lightgray},
        frame=single, 
        frameround=tttt,
        framesep=0.1cm,
        texcl,                                                          
        moredelim=[is][\ttfamily]{<code>}{</code>}, 
        tabsize=4,
	tab=\rightarrowfill,
        xleftmargin=0.25cm,
        xrightmargin=0.25cm,
        showspaces=false,
        showtabs=false,
        columns=fullflexible,
        numberstyle=\tiny,
        keepspaces=true
}

\newcommand{\highlight}[2]{\ifnum\value{lstnumber}>#1\ifnum\value{lstnumber}<#2\color{lightgray}\fi\fi}

\title{Verification of a Smart Contract for a Simple Casino}

\titlerunning{Verification of a Casino}

\author{Mark Utting}{ITEE, The University of Queensland, Australia \and \url{https://researchers.uq.edu.au/researcher/26046}}{m.utting@uq.edu.au}{https://orcid.org/0000-0003-3134-6306}{}

\author{Liam Kent}{ITEE, The University of Queensland, Australia}{l.kent@uq.edu.au}{https://orcid.org/0000-0002-1825-0097}{}

\authorrunning{M. Utting and L. Kent}

\Copyright{Mark Utting and Liam Kent} 



\category{short} 

\relatedversion{} 

\nolinenumbers 

\hideOASIcs 

\EventEditors{John Q. Open and Joan R. Access}
\EventNoEds{2}
\EventLongTitle{42nd Conference on Very Important Topics (CVIT 2016)}
\EventShortTitle{CVIT 2016}
\EventAcronym{CVIT}
\EventYear{2016}
\EventDate{December 24--27, 2016}
\EventLocation{Little Whinging, United Kingdom}
\EventLogo{}
\SeriesVolume{42}
\ArticleNo{23}

\begin{document}

\maketitle


\begin{abstract}
We describe the verification of an existing smart contract for a simple casino application, using the Whiley specification and programming language, with a fully automated verification engine based on Boogie and Z3.
After finding and fixing several specification and code issues in the smart contract,
we are able to verify all the operations of the smart contract.
\end{abstract}

\section{Introduction}

Software updates are standard practice --- even the software in a helicopter on Mars can be updated to fix bugs \cite{mars21}.
But smart contracts on blockchains cannot be updated --- once a smart contract is deployed on a blockchain, it is immutable, by design, so the bugs cannot be fixed.  This makes the correctness of smart contracts a critical issue, that should be addressed with the highest level of verification available.  The gold standard for software verification is machine-checked formal proof of correctness, followed by verified refinement down to the executable machine code or bytecode \cite{Velisarios:2018}.

In this paper, we demonstrate some progress towards the first part of this goal.  We explore the use of an off-the-shelf verification language and tools to verify important safety properties of a smart contract for a simple casino application. The main contribution of our work is to show that modern verification systems that use SMT solvers for back-end reasoning \cite{Utting:boogie2017} are capable of verifying safety properties of simple smart contracts such as this casino example.  We do not address verified refinement down to executable bytecode. 

Section \ref{sec:casino} introduces the Casino example and its existing specification, then Section \ref{sec:verify} describes our use of the Whiley verification language \cite{Pearce:lessons2015} to verify the smart contract.  Section \ref{sec:related} discusses related work and Section \ref{sec:concl} concludes and identifies future work.

\section{The Casino Smart Contract} \label{sec:casino}

\begin{figure}
    \begin{enumerate}
        \item The casino owner may deposit or withdraw money from the casino’s bank as long as the bank’s balance never falls below zero.
        \item As long as no game is in progress, the owner of the casino may make available a new game by tossing a coin and hiding its outcome. The owner must also set a participation cost of choice for the game.
        \item Clauses 1 and 2 are constrained in that as long as a game is in progress, the bank balance may never be less than the sum of the participation cost of the game and its win-out.
        \item The win-out for a game is set to be 80\% of the participating cost.
    \end{enumerate}
    \caption{Excerpt from a legal contract regulating a coin-tossing casino, from Ahrendt \emph{et al.} \cite{Ahrendt:2019}}
    \label{fig:rules}
\end{figure}

Our case study is an Ethereum smart contract \cite{wood2014ethereum} for a simple Casino that follows the rules shown in Fig.~\ref{fig:rules}.
One player at a time can place a wager on whether a hidden coin is heads or tails.
If the player guesses incorrectly, they lose their wager, but if they guess correctly then they are paid 1.8 times their wager.  So the average return to a player would be $(0 + 1.8)/2 = 0.9$ times their wager, giving the casino an expected return of 10\% over the long term.  But interestingly, the actual Solidity contract for this case study \cite{Ahrendt:2019} returns 2.0 times the wager when the player wins, so our casino is actually a non-profit casino!
The casino owner can set the wager amount required to play the game, and can also withdraw or deposit money within certain constraints.

Ahrendt and his co-authors \cite{Ahrendt:2019} wrote a Solidity \cite{solidity} contract\footnote{See \url{https://git.io/fx6cn} --- 219 source lines of code (SLOC) including comments.} to model this casino, then translated that contract into Java with JML \cite{Leavens98jml:a} specifications\footnote{See \url{https://git.io/JONSW} --- 732 SLOC of Java and JML.} and used the KeY verification tool \cite{key-book:2016} to verify two of the methods.  They developed a translation tool called JAVADITY\footnote{See \url{https://github.com/rebiscov/Javadity}.} to translate Solidity contracts into Java, and then added JML specifications to the generated Java code.  KeY has native support for JavaCard, which is a Java derivative that supports a transaction mechanism with rollback of interrupted transactions.  They used these JavaCard features to model the state-reverting exceptions of Solidity, which ensure that the state is left unchanged when errors arise.

They then used KeY to verify that a selection of the methods (\texttt{call\_closeCasino} and \texttt{call\_removeFromPot}) satisfy their contracts and behave correctly when exceptions arise.  The proof of \texttt{call\_removeFromPot} required 22,000 proof steps --- most of these were automated, but 207 interactive proof steps were required \cite{Ahrendt:2019}.

The next section describes how we build on top of this work, to verify the functional behavior of all the methods, using the Whiley verification language \cite{Pearce:lessons2015} with the fully automated back-end prover Boogie \cite{Utting:boogie2017} and the Z3 SMT solver.

\section{Verification with Whiley} \label{sec:verify}

Whiley\footnote{See \url{http://whiley.org}.} is a verification-oriented programming language with the following features:
\begin{itemize}
    \item Python-like syntax that uses indentation to indicate block structure;
    \item static typing with support for records, arrays, union types, and stored functions/methods;
    \item specification features such as \texttt{requires} and \texttt{ensures} clauses for preconditions and postconditions, invariants for loops and records, and \emph{properties}, which are pure boolean functions that are used only within specifications, not in code; 
    \item infinite precision integers to avoid problems with overflow;
    \item call-by-value parameter passing to eliminate aliasing.  Note that even arrays and records are passed by value, and there are no global variables (only constants), so all aliasing is eliminated in order to make verification easier;
    \item support for imperative programming (methods) as well as pure functional programming (functions).
\end{itemize}

We translate the Java+JML version of the Casino smart contract, developed by Ahrendt \emph{et al.}~\cite{Ahrendt:2019}, into Whiley as a proof-of-concept for using Whiley to verify Solidity smart contracts.  Note that they used JavaCard failure semantics to model the transaction-based revert-on-exception control flow of smart contracts, but Whiley does not have this specific failure semantics, so we do not attempt to model this behaviour.  Thus our Whiley model of the smart contract models only the success case, where operations succeed upon valid input to the contract.  We do not model the contract failure cases where a requirement for valid input is not satisfied and the contract state reverts to the state before the invalid call.  Such failure cases could be added to our specification by weakening preconditions and adding if-else statements to handle invalid inputs, but this is outside the scope of this paper.

We make several significant structural changes to the smart contract as we translate it from Java to Whiley.
Firstly, the \texttt{uint256} type is implemented in Whiley as a subrange of the Whiley $int$ type (which is unbounded), with its range defined to be $0$ to $2^{256} - 1$.  This is significantly simpler than the Java version where it was defined as an interface that required re-implementing basic arithmetic operations. This also has the advantage that the back-end prover sees just a subrange of the standard $int$ type, without wrapping, so it can easily use its standard linear arithmetic solver to check that calculations do not overflow or underflow.

Secondly, Whiley promotes the functional and imperative programming styles, and has limited support for object-orientation,\footnote{Whiley does support storing functions or methods within records, and also supports record extension, which is similar to single inheritance, but it does not directly support class-based object-orientation.} so we follow a more functional style than the Java version.  In the Java version the contract as a whole was modelled as a $Casino$ class, but in our Whiley translation, we model the casino state as a $Casino$ record (Fig.~\ref{fig:Casino}) and model each contract operation as a function that takes the $Casino$ record as input and returns a $Casino$ record as output, representing the contract's modified state.  This is illustrated in Fig.~\ref{fig:PlayerWins}, which shows the Solidity, Java+JML and Whiley versions of the \texttt{playerWins} method.

\begin{figure}[htp]
\begin{lstlisting}[language=Whiley,basicstyle=\small]
    public type Casino is {
        Address address,           State state,
        Address operator,         uint256 pot,
        uint256 timeout,          uint256 secretNumber,
        Address player,           Wager wager,
        Message msg,             Block block,
        Transaction tx,          bool destroyed
    }
    where state == BET_PLACED ==> pot + wager.value == address.balance
    where state != BET_PLACED ==> pot == address.balance
    where operator.address != address.address
    where player.address != address.address
    where msg.sender.address != address.address
    where block.coinbase.address != address.address
    where tx.origin.address != address.address
\end{lstlisting}
\caption{The Whiley \texttt{Casino} record with invariant.  This models the whole smart contract state.}
\label{fig:Casino}
\end{figure}

\begin{figure}[t]
\begin{lstlisting}[language=Java,basicstyle=\small]
// SOLIDITY: Player wins and gets back twice his original wager
function playerWins() private {
    pot = pot - wager.value;
    player.transfer(wager.value*2);
    wager.value = 0;
}
\end{lstlisting}
\begin{lstlisting}[language=Java,basicstyle=\small]
/*@ private normal_behavior
    @ requires wager.value.mul(Uint256.TWO).leq(player.balance);
    @ ensures pot.eq(\old(pot.sub(wager.value)));
    @ ensures player.balance.eq(\old(player.balance.sum(wager.value.mul(Uint256.TWO))));
    @ ensures wager.value.eq(Uint256.ZERO);
    @ ensures this.balance.eq(\old(this.balance.sub(wager.value.mul(Uint256.TWO))));
    @ assignable this.pot, wager.value, this.balance, player.balance;
@ also
    @ public exceptional_behavior
    @ requires wager.value.mul(Uint256.TWO).gr(player.balance);
    @ signals (Exception) true;
    @ assignable pot;
    @*/
private void playerWins() throws Exception {
        pot = pot.sub(wager.value);
        player.transfer(this, wager.value.mul(Uint256.TWO));
        wager.value = Uint256.ZERO;
}
\end{lstlisting}
\begin{lstlisting}[language=Whiley,basicstyle=\small]
function playerWins(Casino casino) -> (Casino out)
requires inState(casino, BET_PLACED)
requires casino.wager.value * 2 <= casino.address.balance
requires casino.wager.value * 2 + casino.player.balance < MAX256
ensures out.pot == casino.pot - casino.wager.value
ensures out.player.balance == casino.player.balance + casino.wager.value * 2
ensures out.wager.value == 0
ensures out.address.balance == casino.address.balance - casino.wager.value * 2:
    (Address a1, Address a2) = transfer(casino.player, casino.address, casino.wager.value * 2)
    (casino.player, casino.address, casino.pot, casino.wager.value) = 
        (a1, a2, casino.pot - casino.wager.value, 0)
    return casino
\end{lstlisting}
\caption{Solidity (top) and Java (middle) versions of the \texttt{PlayerWins} method \cite{Ahrendt:2019} compared with our Whiley translation (bottom).}
\label{fig:PlayerWins}
\end{figure}

Thirdly, we strengthen the specifications of the $call\_$ methods (the public smart contract functions), even though some of them did not have specifications in the Java version.  In that work, the authors focussed on verifying just two of the methods, due to the complexity of verifying the JavaCard rollback functionality, but in Whiley we specify all the methods in order to verify them.  However, we do not model rollback functionality.

Other more minor differences between the Java model and the Whiley model include: 
\begin{itemize}
\item Solidity function modifiers that were modelled as function calls in Java and used for run-time validation checks are modelled as Whiley properties which are verified in the function preconditions, as no run-time error-checking behaviour is included in the Whiley translation.  However, the Solidity modifier $payable$ is still modelled in Whiley by a function call, as in the Java version, because its behaviour is more than just run-time error checking and it modifies the contract state.
\item The $State$ and $Coin$ enums are implemented using Whiley's type definitions and constant global variables, as unlike Solidity and Java, Whiley does not have enums. 
\item In the $init$ function which initialises the $Casino$ record state in Whiley, all fields of the $Casino$ record must be initialised, unlike in the Java $Casino$ class's constructor which leaves some fields uninitialised.  We use placeholder values to initialise such fields.
\item The Solidity $send$ function, which was implemented in the Java version, is not implemented in the Whiley translation, as it is not called elsewhere in the contract, and is now deprecated in Solidity.
\item The Solidity $fallback$ function, which is called by default when a call is made to any unknown function, is not implemented in the Whiley translation, as any calls to it can be considered invalid input, and it has an empty body in the Solidity contract.
\end{itemize}

\subsection{Specification Issues and Errors}

The verification of the Whiley specification of the casino smart contract was an interative process.
Typically, when the proof of a particular postcondition reported an error, we investigated that error and found that a precondition needed to be strengthened, or extra information was needed in the invariant, or there was an error in the code of a method or its specifications.  We repeated this process until all issues were fixed, and all methods could be proved correct.  The specification issues that we found during our verification process were all in the methods that had \emph{not} been verified by KeY.  The issues were:

\begin{enumerate}
\item The $decideBet$ method was incorrectly specified in the Java version, saying that the player would get just their original wager returned.  This was inconsistent with the specification of the $playerWins$ method that it calls, which said that twice the wager was returned.  So we changed the $decideBet$ specification to be consistent with the $playerWins$ method and return twice the wager.  This was necessary in order for Whiley to verify these methods.
\item The $removeFromPot$ method in the Java version was implemented incorrectly, with it calling the modifier $payable$ which indicates that money can be sent to the contract with a call to this method. This method is supposed to allow the owner of the contract to withdraw money from the pot, and was specified as such, but including the call to $payable$ meant that money could also be transferred to the casino contract itself.  This inconsistency in the original contract was exposed by one of the Whiley proofs failing, and had to be corrected for the proof to succeed.
\item The $payable$ and $transfer$ methods in the Java version were also implemented and specified incorrectly, checking that the amount of money sent was not less than the \emph{receiver} balance, instead of the \emph{sender} balance.  The Whiley verifier complained that this could result in sender underflow, so we had to correct these inconsistencies before the Whiley proof could succeed.
\end{enumerate}

\subsection{Verification Results}

The final casino contract in Whiley (see Appendix A) consists of 10 type definitions, 6 property definitions (predicates that are used in the specifications), and 24 functions that implement the public and private casino operations, with a total of 360 lines of code (including comments), of which over half (182 lines) are specification lines (properties and \texttt{requires} and \texttt{ensures} clauses).  The number of specification lines would be significantly reduced if we inlined some of the private methods into the public methods, but we decided not to do this, in order to retain the structure of the Java specification.

The whole Whiley contract takes on average 3.6 seconds for the Whiley compiler to parse, typecheck and translate to the Boogie intermediate language, then 4.1 seconds to verify using Boogie v2.8.26 and Z3 v4.8.10, on a Windows 10 laptop with i7-7820HQ CPU at 2.9GHz.  

The main benefit of this verification approach is that the prover is fully automatic, once the specifications have been written.
In general, because Whiley verification is undecidable, verification attempts may report timeout errors, meaning that a function was not able to be verified in a given time.  But for this particular example, the prover could easily verify \emph{all} the functions of the contract.
Since smart contracts usually eschew general loops and complex data structures, and typically just contain conditionals and assignments, there is reason to expect that many smart contracts may be able to be fully verified in this way.

\section{Related Work} \label{sec:related}

The most closely related work is Ahrendt \emph{et al.}~\cite{Ahrendt:2019}, which we have built on in this paper.

Surveys of verification approaches for smart contracts \cite{harz:survey:2018,Almakhour:survey:2020} and the more recent systematic review of smart contract construction and execution techniques by Hu \emph{et al.}~\cite{hu:survey:2021} show that tool support for formal verification of smart contracts is an increasingly active research topic, but the tools are still in their infancy.
Most of the verification approaches work at the EVM bytecode level \cite{Bhargavan:2016,Amani:Isabelle:2018}, since this has a clearer formal semantics than Solidity \cite{Hirai:EVM:2017}.  Some researchers have proposed translating Solidity into new intermediate-level languages with formal semantics \cite{Yang_2020}, thus enabling reasoning about the contracts using established provers such as Coq (an interactive prover).

Our approach has been to investigate whether an existing verification language with a fully automatic prover (e.g. Whiley with Boogie and Z3 as the proof engine) is suitable for verifying smart contracts.  Our case study indicates that it can be.
However, we do not address here the issue of having some verified connection between the Whiley contract and the low-level EVM bytecode that would implement that contract.

\section{Conclusions and Future Work} \label{sec:concl}

We have shown that off-the-shelf software verification tools can usefully be applied to verifying the functional behavior of smart contracts.
Our approach was able to benefit from fully automatic verification tools, but requires expressing the contract specifications and code in a suitable language that supports formal verification.  
Here we used Whiley, but we expect that similar results could be obtained using other verification systems that are based on SMT solvers, such as Dafny \cite{leino:dafny:2010}, which also uses Boogie and Z3 as back-end provers.

One limitation of our approach is that we verified only the normal-execution code path, whereas the KeY approach also verified that the failure cases correctly rolled-back to the original state. 
But the main benefit of our approach is that the verification was automatic, so did not require any interactive proving skills, and it was able to easily verify all the functions of the contract rather than just the main two functions.
Our verification approach also discovered three specification errors in the JML specifications, and added value by strengthening method preconditions in order to verify that 256-bit integer calculations never overflow or underflow.

Useful future work would be to investigate the second phase of verification mentioned in the introduction --- verifying the translation of a high-level contract such as our Whiley contract into the low-level EVM bytecode that is actually executed.  This would require a formal semantics of EVM (e.g. \cite{Hirai:EVM:2017}) and then a significant compiler verification project.



\bibliography{../bibtex/veri-smart-refs}

\appendix
\newpage
\section{Whiley version of the Casino Contract} \label{sec:listing}

See GitHub
\url{https://github.com/uqcyber/VeriSmart/tree/main/casino}
for full code.

\begin{lstlisting}[language=Whiley,basicstyle=\footnotesize]
import std::math

// public final int MAX256 = math::pow(2, 256)
public final int MAX256 = 
  115792089237316195423570985008687907853269984665640564039457584007913129639936
public type uint256 is (int x) where x >= 0 && x < MAX256 
 
/* TODO: add this.  It was not used in base implementation.
public function keccak256(uint256 in) -> (uint256 out):
    return 0 
*/

// 2**160 - solidity addresses are 160-bit integers
public type uint160 is (int x) where x >= 0 &&
    x < 1461501637330902918203684832716283019655932542976

public type Address is {
    uint160 address, // should be immutable, unique
    uint256 balance
}

public type Block is {
    Address coinbase,
    uint256 difficulty,
    uint256 gaslimit, 
    uint256 number,
    uint256 timestamp
}

public type Message is {
    Address sender,
    uint256 gas,
    uint160 data,
    uint256 value
}

public type Transaction is {
    uint256 gasprice,
    Address origin
}

public type Coin is (int x) where x == 0 || x == 1
public final Coin HEADS = 0
public final Coin TAILS = 1

public type State is (int x) where x == 0 || x == 1 || x == 2
public final State IDLE = 0
public final State GAME_AVAILABLE = 1
public final State BET_PLACED = 2 

public type Wager is {
    uint256 value,
    Coin guess,
    uint256 timestamp
}

// main contract state class with all global variables
public type Casino is {
    Address address,
    State state,
    Address operator,
    uint256 pot,
    uint256 timeout,
    uint256 secretNumber,
    Address player,
    Wager wager,
    Message msg,
    Block block,
    Transaction tx,
    bool destroyed
}
where state == BET_PLACED ==> pot + wager.value == address.balance
where state != BET_PLACED ==> pot == address.balance
where operator.address != address.address
where player.address != address.address
where msg.sender.address != address.address
where block.coinbase.address != address.address
where tx.origin.address != address.address


// executed by address receiving the money which withdraws 'price' from the 'sender' address
public function transfer(Address receiver, Address sender, uint256 price) -> 
    (Address receiver_update, Address sender_update)
requires price <= sender.balance // different in Java, incorrectly
requires sender.address == receiver.address ==> sender.balance == receiver.balance
requires receiver.balance + price < MAX256
ensures receiver_update.address == receiver.address
ensures sender_update.address == sender.address 
ensures sender.address != receiver.address ==> (receiver_update.balance == 
    receiver.balance + price && sender_update.balance == sender.balance - price)
ensures sender.address == receiver.address ==> (sender_update.balance == 
    receiver_update.balance && sender_update.balance == sender.balance):
    if sender.address != receiver.address:
        sender.balance = sender.balance - price
        receiver.balance = receiver.balance + price
    return (receiver, sender)

public function payable(Address receiver, Message msg) -> 
    (Address receiver_update, Address sender_update)
requires msg.value <= msg.sender.balance // different in Java, incorrectly
requires receiver.address == msg.sender.address ==> msg.sender.balance == receiver.balance
requires receiver.balance + msg.value < MAX256
ensures receiver_update.address == receiver.address
ensures sender_update.address == msg.sender.address
ensures receiver.address != msg.sender.address ==> (receiver_update.balance == receiver.balance
     + msg.value && sender_update.balance == msg.sender.balance - msg.value)
ensures receiver.address == msg.sender.address ==> (receiver_update.balance == 
        sender_update.balance && receiver.balance == receiver_update.balance):
    (receiver, msg.sender) = transfer(receiver, msg.sender, msg.value)
    return (receiver, msg.sender)

function init(uint160 contract_address, Message msg, Block block, Transaction tx)->(Casino out)
requires contract_address != 0
requires msg.sender.address != contract_address
requires block.coinbase.address != contract_address
requires tx.origin.address != contract_address:
    return {
        address: {address: contract_address, balance: 0},
        state: IDLE,
        operator: msg.sender,
        pot: 0,
        timeout: 30,
        secretNumber: 0, // default 
        player: {address: 0, balance: 0}, // default 
        wager: {value: 0, guess: HEADS, timestamp: 0}, // default 
        msg: msg,
        block: block, 
        tx: tx, 
        destroyed: false
    }

public property costs(Casino casino, uint256 value)
where casino.msg.value == value

public property byOperator(Casino casino)
where casino.msg.sender == casino.operator

public property noActiveBet(Casino casino)
where casino.state == IDLE || casino.state == GAME_AVAILABLE

public property inState(Casino casino, State state)
where casino.state == state

public property validCall(Casino casino, Message msg, Block block, Transaction tx)
where msg.sender.address != casino.address.address
where block.coinbase.address != casino.address.address
where tx.origin.address != casino.address.address

function updateBlockChainVariables(Casino casino, Message msg, Block block, 
            Transaction tx) -> (Casino out)
requires validCall(casino, msg, block, tx)
ensures casino.address == out.address
ensures casino.state == out.state
ensures casino.operator == out.operator
ensures casino.pot == out.pot
ensures casino.timeout == out.timeout
ensures casino.secretNumber == out.secretNumber
ensures casino.player == out.player
ensures casino.wager == out.wager
ensures casino.destroyed == out.destroyed
ensures out.msg == msg
ensures out.block == block
ensures out.tx == tx:
    casino.msg = msg
    casino.block = block
    casino.tx = tx
    return casino


public function call_addToPot(Casino casino, uint256 value, Message msg, 
             Block block, Transaction tx) -> (Casino out)
requires validCall(casino, msg, block, tx)
requires msg.sender == casino.operator
requires msg.value == value && value > 0
requires value <= casino.operator.balance
requires casino.address.balance + value < MAX256
ensures out.pot == casino.pot + value:
    casino = updateBlockChainVariables(casino, msg, block, tx)
    casino = addToPot(casino, value)
    return casino

function addToPot(Casino casino, uint256 value) -> (Casino out)
requires byOperator(casino)
requires costs(casino, value)
requires value > 0
requires value <= casino.operator.balance
requires casino.address.balance + value < MAX256
ensures out.pot == casino.pot + value:
    (Address a1, Address a2) = payable(casino.address, casino.msg)
    casino.address, casino.msg.sender, casino.pot = a1, a2, casino.pot + value
    return casino


public function call_createGame(Casino casino, uint256 secretNumber, Message msg, 
            Block block, Transaction tx) -> (Casino out)
requires validCall(casino, msg, block, tx)
requires inState(casino, IDLE)
requires msg.sender == casino.operator:
    casino = updateBlockChainVariables(casino, msg, block, tx)
    casino = createGame(casino, secretNumber)
    return casino

function createGame(Casino casino, uint256 secretNumber) -> (Casino out)
requires byOperator(casino)
requires inState(casino, IDLE)
ensures out.secretNumber == secretNumber
ensures out.state == GAME_AVAILABLE:
    casino.secretNumber = secretNumber
    casino.state = GAME_AVAILABLE
    return casino


public function call_placeBet(Casino casino, uint256 value, Coin guess, Message msg,
            Block block, Transaction tx) -> (Casino out)
requires validCall(casino, msg, block, tx)
requires inState(casino, GAME_AVAILABLE)
requires msg.sender.address != casino.operator.address
requires value > 0 && value <= casino.pot
requires msg.value == value
requires value <= msg.sender.balance
requires casino.address.balance + value < MAX256
ensures out.state == BET_PLACED
ensures out.player.address == msg.sender.address
ensures out.wager.value == value
ensures out.wager.guess == guess
ensures out.wager.timestamp == block.timestamp:
    casino = updateBlockChainVariables(casino, msg, block, tx)
    casino = placeBet(casino, value, guess)
    return casino

function placeBet(Casino casino, uint256 value, Coin guess) -> (Casino out)
requires costs(casino, value)
requires inState(casino, GAME_AVAILABLE)
requires casino.msg.sender.address != casino.operator.address
requires value > 0 && value <= casino.pot
requires value <= casino.msg.sender.balance
requires casino.address.balance + value < MAX256
ensures out.state == BET_PLACED
ensures out.player.address == casino.msg.sender.address
ensures out.wager.value == value
ensures out.wager.guess == guess
ensures out.wager.timestamp == casino.block.timestamp:
    (Address a1, Address a2) = payable(casino.address, casino.msg)
    (casino.address, casino.msg.sender, casino.state, casino.player, casino.wager) = 
        (a1, a2, BET_PLACED, casino.msg.sender, {
		value: value,
		guess: guess,
		timestamp: casino.block.timestamp
         })
    return casino


public function call_decideBet(Casino casino, uint256 publicNumber, Message msg, 
             Block block, Transaction tx) -> (Casino out)
requires validCall(casino, msg, block, tx)
requires inState(casino, BET_PLACED)
requires msg.sender == casino.operator
requires (publicNumber % 2 == 0 <==> casino.wager.guess == HEADS) ==>
    (casino.wager.value * 2 <= casino.address.balance && 
     casino.wager.value * 2 + casino.player.balance < MAX256)
//requires casino.secretNumber == keccak256(publicNumber) // ignore for now
ensures (publicNumber % 2 == 0 <==> casino.wager.guess == HEADS) ==>
    (out.pot == casino.pot - casino.wager.value && 
     out.address.balance == casino.address.balance - casino.wager.value * 2)
ensures !(publicNumber % 2 == 0 <==> casino.wager.guess == HEADS) ==> 
    out.pot == casino.pot + casino.wager.value
ensures out.wager.value == 0
ensures out.state == IDLE:
    casino = updateBlockChainVariables(casino, msg, block, tx)
    casino = decideBet(casino, publicNumber)
    return casino

function decideBet(Casino casino, uint256 publicNumber) -> (Casino out)
requires byOperator(casino)
requires inState(casino, BET_PLACED)
requires (publicNumber % 2 == 0 <==> casino.wager.guess == HEADS) ==>
    (casino.wager.value * 2 <= casino.address.balance && 
     casino.wager.value * 2 + casino.player.balance < MAX256)
//requires casino.secretNumber == keccak256(publicNumber) // ignore for now
ensures (publicNumber % 2 == 0 <==> casino.wager.guess == HEADS) ==>
    (out.pot == casino.pot - casino.wager.value &&
     out.address.balance == casino.address.balance - casino.wager.value * 2)
ensures !(publicNumber % 2 == 0 <==> casino.wager.guess == HEADS) ==> 
    out.pot == casino.pot + casino.wager.value
ensures out.wager.value == 0
ensures out.state == IDLE:
    Coin secret
    if (publicNumber % 2 == 0):
        secret = HEADS
    else:
        secret = TAILS
    if (secret == casino.wager.guess):
        casino = playerWins(casino)
    else:
        casino = operatorWins(casino)
    casino.state = IDLE
    return casino

function playerWins(Casino casino) -> (Casino out) 
requires inState(casino, BET_PLACED)
requires casino.wager.value * 2 <= casino.address.balance
requires casino.wager.value * 2 + casino.player.balance < MAX256
ensures out.pot == casino.pot - casino.wager.value
ensures out.player.balance == casino.player.balance + casino.wager.value * 2
ensures out.wager.value == 0
ensures out.address.balance == casino.address.balance - casino.wager.value * 2:
    (Address a1, Address a2) = transfer(casino.player, casino.address, casino.wager.value * 2)
    (casino.player, casino.address, casino.pot, casino.wager.value) = 
        (a1, a2, casino.pot - casino.wager.value, 0)
    return casino

function operatorWins(Casino casino) -> (Casino out)
requires inState(casino, BET_PLACED)
ensures out.pot == casino.pot + casino.wager.value
ensures out.wager.value == 0:
    (casino.pot, casino.wager.value) = (casino.pot + casino.wager.value, 0)
    return casino

// The following functions are omitted from this paper to save space.
// See the GitHub repository for full details...
public function call_removeFromPot(Casino casino, ...) -> (Casino out)
public function call_timeoutBet(Casino casino, ...) -> (Casino out)
public function call_updateTimeout(Casino casino, ...) -> (Casino out)
public function call_closeCasino(Casino casino, ...) -> (Casino out)
function selfdestruct(Casino casino, Address a) -> (Casino out, Address receiver)
\end{lstlisting}

\end{document}